\documentclass[traditabstract,paper]{aa}
\usepackage[bookmarks,bookmarksnumbered,colorlinks=true, citecolor=blue,anchorcolor=blue, urlcolor=cyan, breaklinks=true, menucolor=black, linktocpage=true]{hyperref}
\usepackage{natbib}
\usepackage{graphicx}
\usepackage{multirow}
\usepackage{latexsym}
\usepackage{txfonts}
\begin{document}
\title{Magnetic field emergence in mesogranular-sized exploding granules observed with $\textsc{Sunrise}$/IMaX data}
\subtitle{}

\author{J. Palacios\inst{1}, J. Blanco Rodr{\'{\i}}guez\inst{1},  S. Vargas Dom\'inguez\inst{2}, V. Domingo\inst{1}, V. Mart{\'{\i}}nez Pillet\inst{3},  J.~A.~Bonet\inst{3}, L.~R.~Bellot~Rubio\inst{4},  J. C. del Toro Iniesta\inst{4}, S. K. Solanki\inst{5,8}, P. Barthol\inst{5}, A. Gandorfer\inst{5}, \\
T. Berkefeld\inst{6}, W. Schmidt\inst{6}, M. Kn\"olker\inst{7}
}
\offprints{J. Palacios Hern\'andez, \email{judith.palacios@uv.es}
}

\institute{ Image Processing Laboratory, University of Valencia, P.O. Box: 22085, E-46980 Paterna, Valencia, Spain  \and Departamento de F\'isica, Universidad de Los Andes, A.A. 4976, Bogot\'a, Colombia \and Instituto de Astrof{\'{\i}}sica de Canarias, E-38205 La Laguna, Tenerife, Spain \and Instituto de Astrof{\'{\i}}sica de Andaluc{\'{\i}}a, Apdo. de Correos 3004, 18080, Granada, Spain \and Max Planck Institut f\"ur Sonnensystemforschung, Max Planck Strasse 2, Katlenburg-Lindau, 37191, Germany \and Kiepenheuer-Institut f\"ur Sonnenphysik, Sch\"oneckstr. 6, 79104 Freiburg, Germany \and High Altitude Observatory, National Center for Atmospheric Research, P.O. Box 3000, Boulder, CO 80307-3000, USA \and School of Space Research, Kyung Hee University, Yongin, Gyeonggi, 446-701, Republic of Korea}


\date{Received Aug 23, 2011; accepted Oct 18 2011}

\def\farcs{\hbox{$.\!\!^{\prime\prime}$}}
\def\arcsec{\hbox{$^{\prime\prime}$}}
\def\kms{{${\rm km}\:{\rm s}^{-1}\:\!\!$}}
\def\dens{\mbox{${\rm km}^{-2}\:{\rm s}^{-1}\:\!$}}

\abstract 
{
{We report on magnetic field emergences covering significant areas of exploding granules.}
{The balloon-borne mission $\textsc{\large Sunrise}$ provided high spatial and temporal resolution images of the solar photosphere. Continuum images, longitudinal and transverse magnetic field maps and Do\-ppler\-grams obtained by IMaX onboard $\textsc{\large Sunrise}$ are analyzed by Local Correlation Traking (LCT), divergence calculation and time slices, Stokes inversions and numerical simulations are also employed.}
{We characterize two mesogranular-scale exploding granules where $\sim$ 10$^{18}$ Mx of magnetic flux emerges. The emergence of weak unipolar longitudinal fields ($\sim$100 G) start with a single visible magnetic polarity, occupying their respective gra\-nules' top and following  the granular splitting. After a while, mixed polarities start appearing, concentrated in downflow lanes. The events last around 20~min. LCT analyses confirm mesogranular scale expansion, displaying a similar pattern for all the physical properties, and divergence centers match between all of them. We found a similar behaviour with the emergence events in a numerical MHD simulation. Granule expansion velocities are around 1~\kms~~while magnetic patches expand at 0.65~\kms~. One of the analyzed events evidences the emergence of a loop-like structure.}
{Advection of the emerging magnetic flux features is dominated by convective motion resulting from the exploding granule due to the magnetic field frozen in the granular plasma. Intensification of the magnetic field occurs in the intergranular lanes, probably because of being directed by the downflowing plasma.}
}
\keywords{Sun: surface magnetism -- Sun: granulation -- Sun: photosphere  -- Techniques: polarimetric}

\authorrunning{Palacios et al.}
\titlerunning{Magnetic field emergence in exploding granules with $\textsc{\large Sunrise}$/IMaX data}
\maketitle
\section{Introduction}
\label{S:1}
The $\textsc{\large Sunrise}$ balloon-borne mission \citep{Barthol2011} was launched in June 2009. Having a 1-m aperture telescope, this state-of-the-art mission provided high resolution images during its nearly 6 day observing campaign. At a cruise altitude of around 36000 meters, the mission eluded 99\% of the perturbations induced by the Earth's atmosphere \citep{Solanki2010}. The Imaging Magnetograph eXperiment IMaX \citep{MartinezPillet2011} onboard $\textsc{\large Sunrise}$ acquired longitudinal and transverse magnetograms from full Stokes measurements of the Fe $\textsc{i}$ at 5250~\AA~line, among other physical quantities.

Although mesogranulation, an intermediate scale between granulation and supergranulation \citep[3-10 Mm and lifetimes between 3-10 hours,][]{November1981}, has been thoroughly studied, the question whether it is a natural convection scale remains unanswered \citep{Cattaneo2001,Yelles2011,Ploner2000}. This scale often manifests itself when temporal series of about 20~min to 1~hour are smoothed or approached through correlation tracking. Mesogranulation is sometimes identified as patches of po\-si\-tive divergence. \citet{Matloch2010} concluded that these patterns do not exhibit intrinsic size or time scales. This particular convection scale can indicate the relation to exploding granules and their fragmenting granule families. \citet{Bonet2005} performed a deep analysis of the granulation scales around a sunspot, focusing on the trees of splitting granules. \citet{Roudier2004} matched families of fragmenting granules with prominent divergences. Passively advected tracers seem to coincide with these families. Mesogranules are considered as structures likely related to exploding granules, whose splitting is preceded by the formation of a dark hub due to buoyancy braking \citep[i.e.,][]{Massaguer1980}. This effect caused by com\-pre\-ssibi\-lity of the gas \citep{Spruit1990} takes place when the mass excess in the center of the granule reduces the upward velocity until its collapse. If energy losses cannot be compensated, the center of the granule cools down, and the dark core emerges. When an incompressible plasma is considered, exploding gra\-nu\-les do not develop, and mesogranulation arises as interaction between granules \citep{Cattaneo2001}. Compressibility and radiative cooling  seem to be the factors which mark the maximum size for a granule. \\
The relationship between mesogranulation and vertical magnetic fields, as network, has been analyzed by a number of authors. It is generally believed that the magnetic network is pulled out to the supergranular boundaries. \citet{deWijn2009} found trees of fragmenting granules devoid of magnetic elements, as they are dragged to the boundaries of the supergranule. \citet{Cerdena2003} concluded that mesogranular structures in the network are smaller than those in the internetwork. Mesogranulation has been recently studied and related to the distance to the network by \citet{Yelles2011}. 

In this work we characterize two events of magnetic flux emergence within mesogranular-sized granular cells. These emergences, unipolar at the beginning, drift to the intergranular lane where the magnetic field intensifies. In or very close to the lane opposite pola\-ri\-ties appear as well. Transverse field is almost negligible over the granules' top, except when some opposite polarity becomes apparent and consequently the feature shows a loop topology, as described by \citet{Centeno2007}, \citet{Martinez2009} and \citet{Gomory2010}. These emergences seem to be ca\-rried passively by the horizontal flow of the ex\-plo\-ding gra\-nule.
Previously, magnetic field emergences over gra\-nules were found by \citet{Pontieu2002, Orozco2008}. Numerical si\-mu\-la\-tions  by \citet{Tortosa2009} showed large abnormal gra\-nules lasting 15 minutes and containing a magnetic field in the form of magnetic loops, as the ones presented here. We compare our detected emergence events with simulated data by \citet{Cheung2008} although the spatial scale of the latter co\-rres\-ponds to supergranular size.
\section{Observations and data processing}
\label{S:2}
\begin{figure*}
\centering
\includegraphics[width=0.49\linewidth]{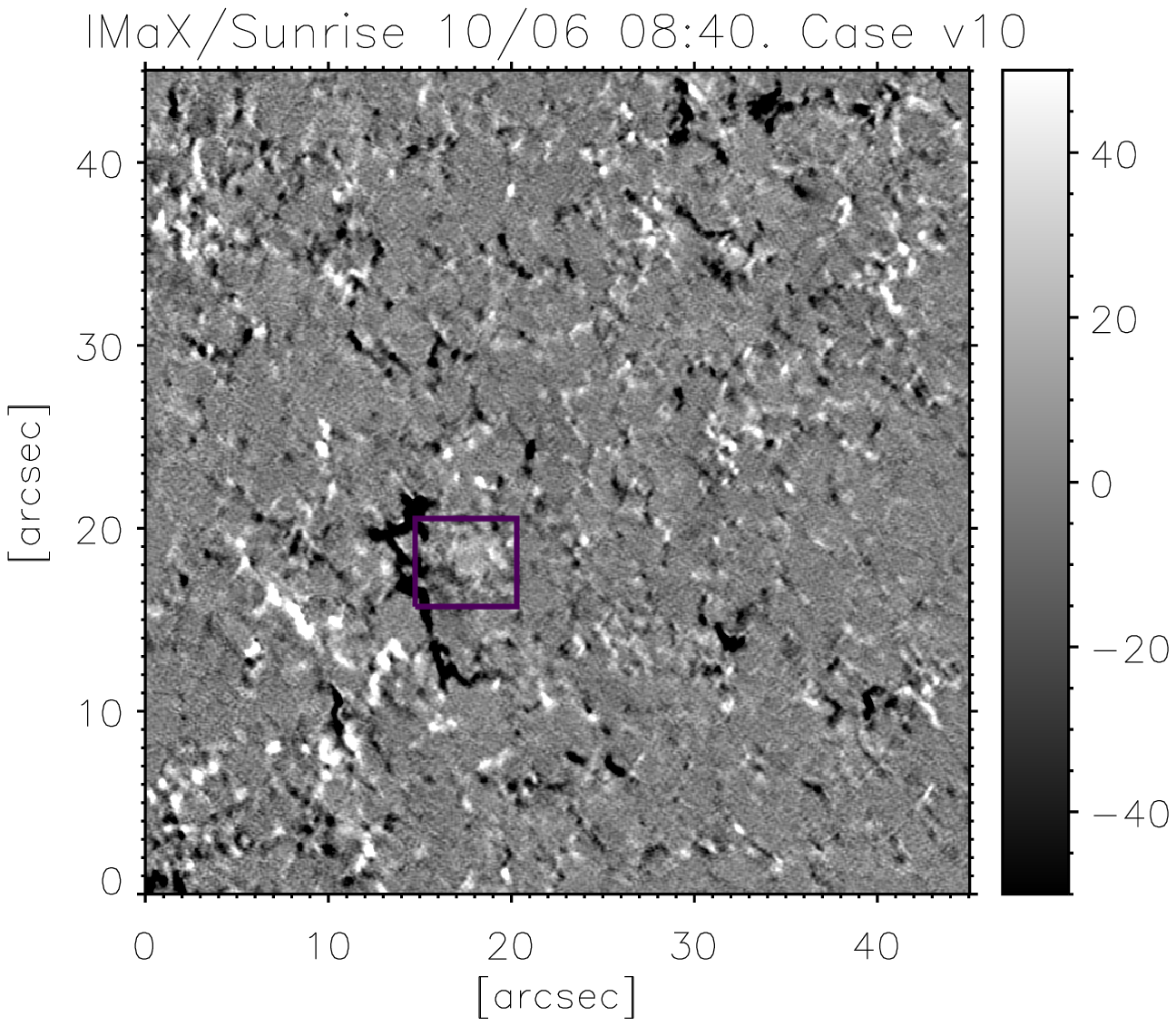}
\includegraphics[width=0.49\linewidth]{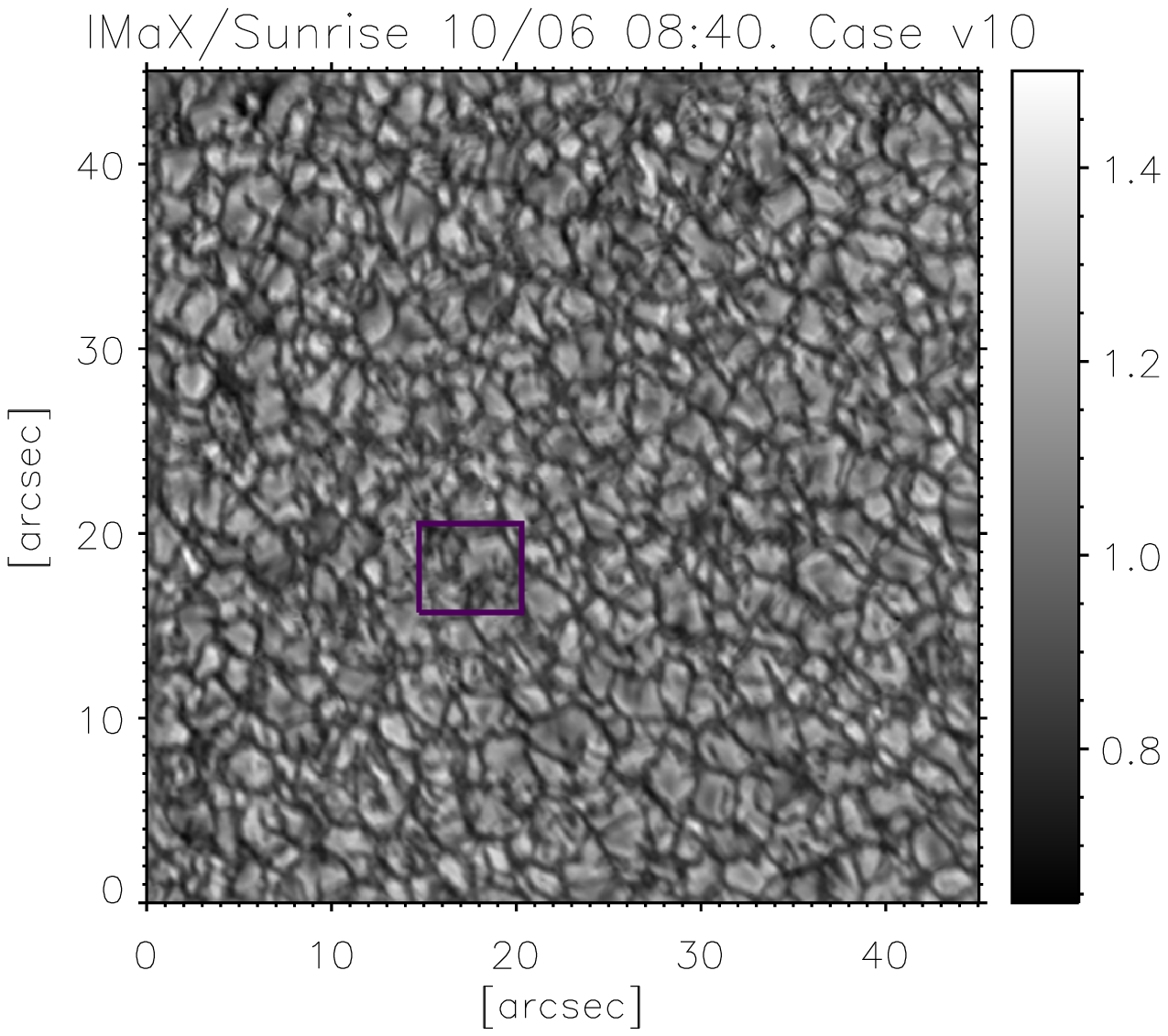}
\includegraphics[width=0.49\linewidth]{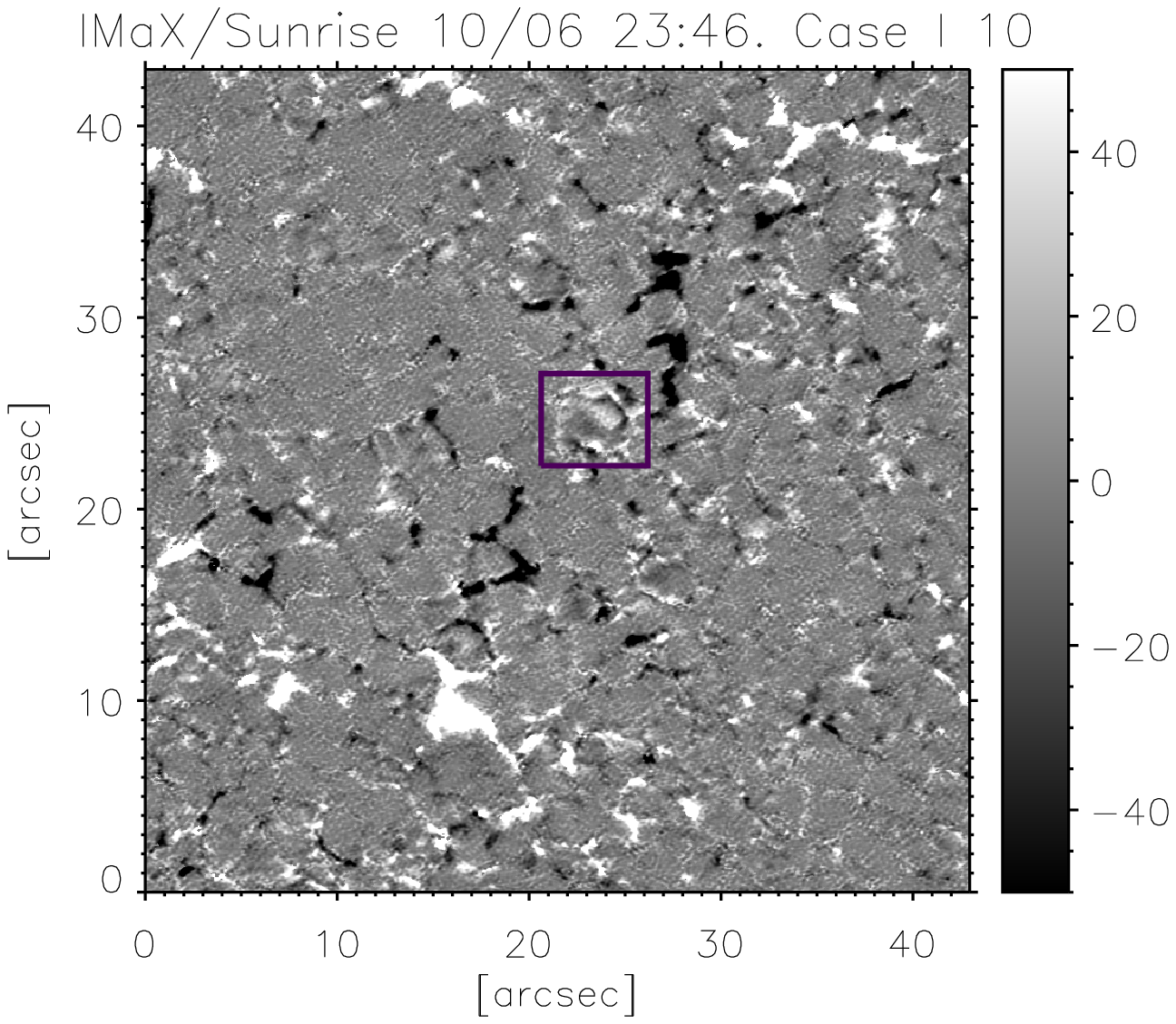}
\includegraphics[width=0.49\linewidth]{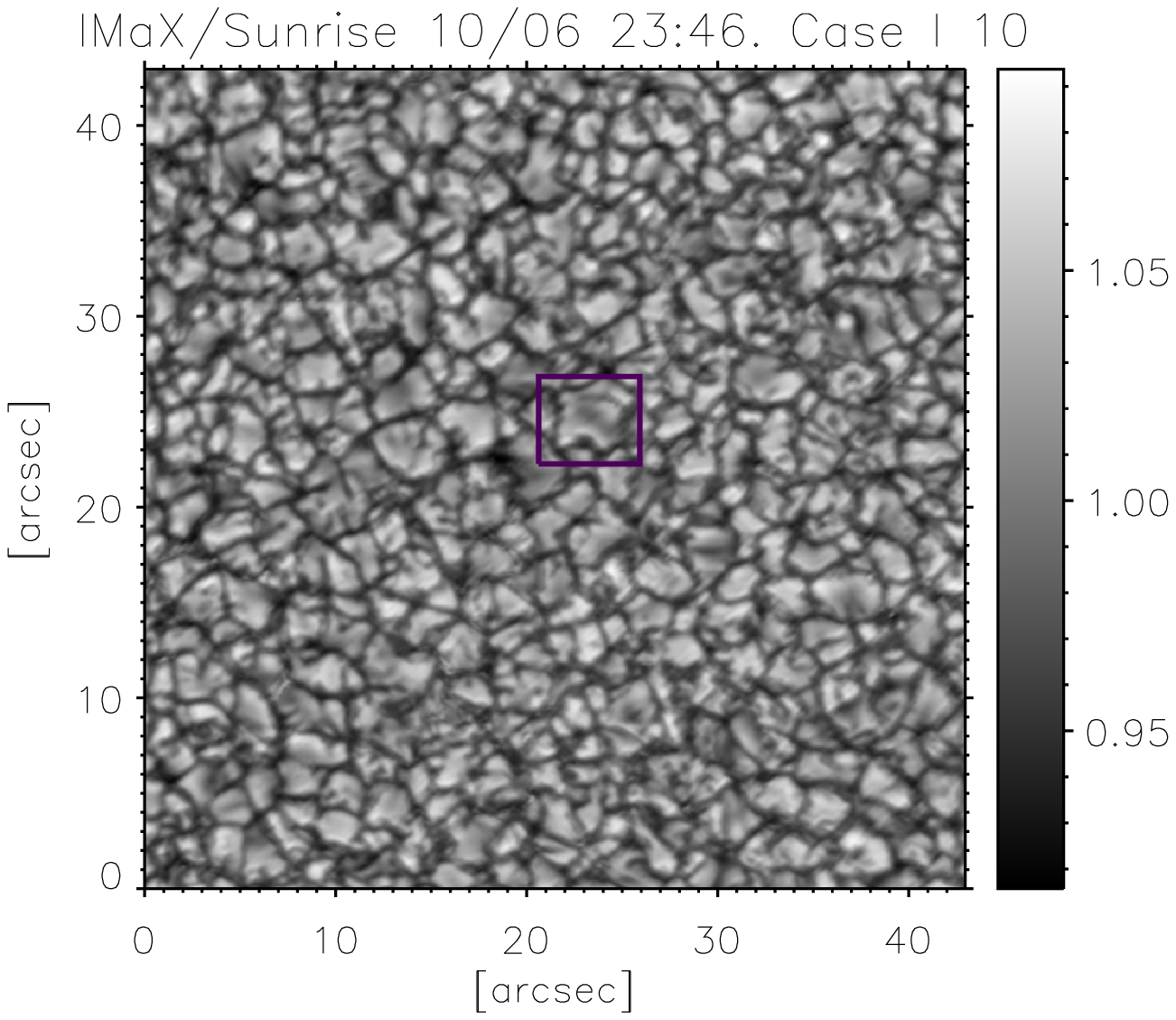}
\caption{Full field-of-view longitudinal magnetic field maps ({\it left}) and continuum images ({\it right}) for the $v10$ event ({\it top row}) and the $\ell10$ event ({\it lower row}). The dark purple squared boxes mark the exploding granules and their corresponding magnetic emergence. The box dimensions are 6\farcs9 $\times$ 6\farcs9. Longitudinal magnetograms are saturated to $\pm$ 50 G.}
\label{context}
\end{figure*}
The instrument IMaX was designed to observe the solar surface in polarized light at very high spatial resolution. IMaX achieved a spectral resolution of 85~m\AA~and a spatial resolution of 0\farcs15-0\farcs18, observing at different wavelengths in the Fe $\textsc{i}$ 5250~\AA~ line. 
The res\-to\-red data present an effective FOV of 45\arcsec$\times$45\arcsec, with a spatial sampling of 0\farcs055. IMaX has several Observing Modes, two of which were used in this study: V5-6 and  L12-2. With six accumulations, the V5-6 Mode includes full Stokes ima\-ging polarimetry at four wavelengths $\pm$ 40, $\pm$ 80 m\AA~from line center and a continuum point at +227 m\AA, with 33~s cadence. Also the L12-2 Mode was employed, where Stokes $I$ and $V$ are observed at twelve wavelengths, from -192.5 to +192.5 m\AA~ in steps of 35 m\AA, and two accumulations, with 29~s cadence. The noise level is below 10$^{-3}$$I_{c}$ in both Modes.\\
The data used in this work were collected by IMaX/$\textsc{\large Sunrise}$ close to the solar disk center and consist of two series in two different Modes, acquired on 2009, June 10. Physical quantities are obtained by different means. The first series (hereafter referred to as emergence event $\it{v10}$) is obtained in IMaX V5-6~Mode, from UT 08:29 to 08:50 ($\sim$21 min). In this case, longitudinal magnetograms are obtained by weak-field approximation \citep{Landi1992}. Transverse magnetograms  and Dopplergram calibrations are used as described in \citet{MartinezPillet2011}, along with the description of the IMaX Observing Modes and image restoration process. The second data series (emergence event $\ell \it{10}$) is acquired in the Longitudinal Mode L12-2, from UT 23:42 to 23:59 ($\sim$18 min). In the $\ell \it{10}$ case, line-of-sight magnetic field strength is obtained by $\bf{1)}$ the center-of-gravity (COG) method \citep[][]{Semel1967,Rees1979} and $\bf{2)}$ SIR inversion \citep{Ruiz1992}. Line-of-sight (LOS) velocities are obtained  from the SIR inversion. All data are subsonic filtered, hence $\it{p-mode}$ removal is applied, following the space-time filtering by \citet{Title1989}. In Fig.~\ref{context}, the full field-of-view images are shown. The dark purple squared boxes mark the exploding granules and their associated magnetic emergence.
\begin{figure*}
\centering
\includegraphics[angle=-90,width=0.95\linewidth]{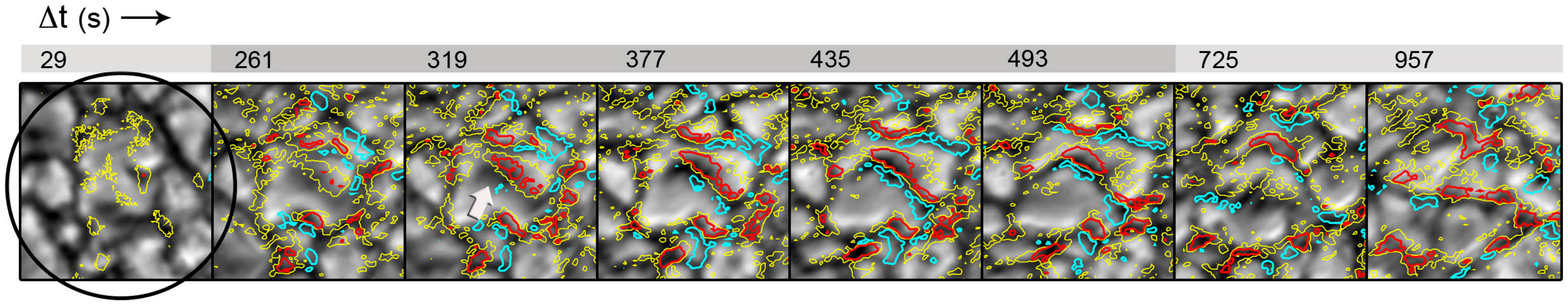}
\caption{Sequence of continuum images for emergence event $\ell \it{10}$. Each image is overlaid with the corresponding contours of longitudinal magnetic field (15 G in yellow; 45 G in red, and blue outlines -45 G). Corresponding times are labelled for each frame in the top bar. The FOV is cropped to a size of 6\farcs9 $\times$ 6\farcs9. The black circumference presents a diameter of 8\farcs3 (6 Mm). A white arrow in the third frame points to the buoyancy braking darkening (see text for details). The corresponding movie for the $\ell \it{10}$ emergence is provided as electronic material.}
\label{contourl122}
\end{figure*}
\section{Analysis}
\label{S:3}
\subsection{Host exploding granules}
\label{SS:1}
Two emergence events ($\it{v10}$, $\ell \it{10}$) are followed until the host exploding granules lose their identity and their fragments develop into three and four granules, respectively. Unfortunately neither of the two time series covers the full development and decay of the fragmenting granules. The granule hosting the magnetic emergence $\it{v10}$ presents a large spatial scale. An image of the granule is chosen before it splits into three gra\-nules. With this particular frame we produce a binary mask via thresholding (which se\-pa\-rates gra\-nules from inter\-gra\-nu\-lar lanes) to estimate granules size. Areas are estimated and the equivalent radius is defined considering that areas correspond to circular granules. Hence histograms of the granular area and the equivalent granule radii are computed (not shown). The host granule equivalent radius and area lies into the last bins of the histograms, since this granule presents a large diameter at the moment when the dark hub appears, being 2300 km ($>$3\arcsec). \citet{Spruit1990} and re\-fe\-rences therein state that the maximum size of granules before the onset of buoyancy braking  is around 3\arcsec. We use the aforementioned binary mask for each frame to compute the number of granules with equivalent diameter larger than 3\arcsec, finding only around 1\%, i.e, 3 granules per each IMaX frame (FOV of 45\arcsec$\times$45\arcsec). We calculate the average granule coverage for the whole FOV of the $\it{v10}$ series and the corresponding equivalent diameter is around 1\farcs22 for regular granules. Besides, in this work we found that the occu\-rrence rate of granules with diameter larger than 3\arcsec~is approximately 5.7 $\times$ 10$^{-11}$ \dens, yielding a similar value as in \citet{Title1989}, where the occu\-rrence rate of exploding granules is found to be 7.7 $\times$ 10$^{-11}$ \dens. \\

The expansion velocity of the host granules is estimated by measuring the area and calculating the equivalent radius for every frame in both series. Then, velocity is calculated by least-squares fitting to the equivalent radius growth rate. The expansion velocity is $0.94\pm0.02$~\kms~for $\ell \it{10}$ and $0.95\pm0.02$~\kms~for $\it{v10}$. Expansion velocities of the magnetic patch covering the ex\-plo\-ding granules are also estimated. Results are $0.61\pm0.04$~\kms~for $\ell \it{10}$ and $0.68\pm0.03$~\kms~for $\it{v10}$. Considering a thres\-hold of 15~G for reconstructed images \citep[see][]{MartinezPillet2011} we measure the area and calculate the equivalent radius for the magnetic patches. 
\vspace{-0.3cm}
\subsection{Associated magnetic field emergences}
\label{SS:2}
\begin{figure}
\centering
\includegraphics[angle=-90,width=0.9\linewidth]{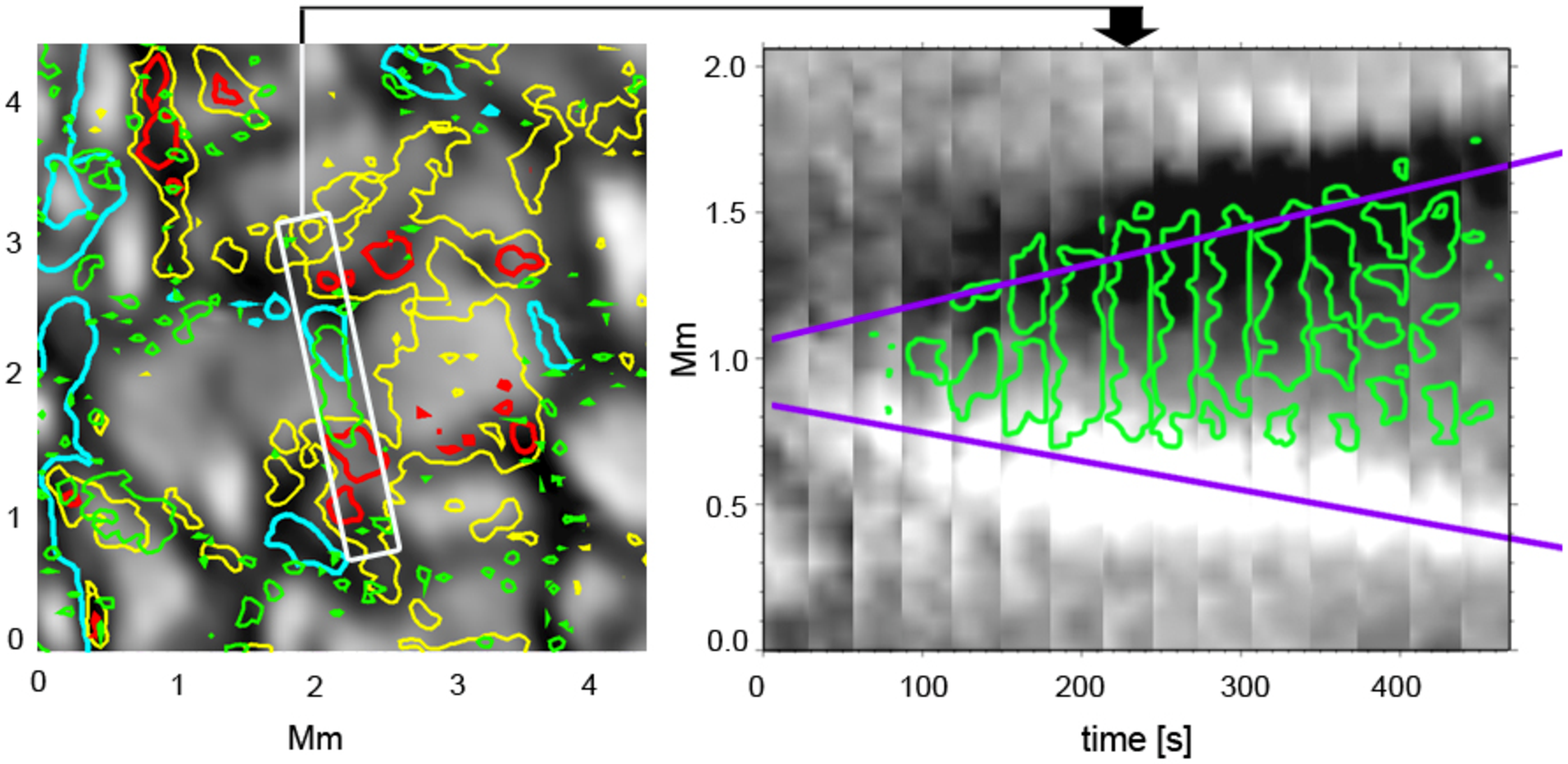}
\caption{Emergence event $\it{v10}$. $\it{Left}$: Continuum image with contours of transverse magnetic field in green ($>$ 180~G) and yellow, blue and red contours as in Fig.~\ref{contourl122}. A region showing the emergence of a loop is outlined with a white box. $Right$: Corresponding time slices of the magnetic loop emergence. Background image displays the longitudinal field with positive/negative polarities in white/black. Superimposed in green is the transverse field. Reference lines in purple mark the footpoint separation. Time in this panel is counted from the start of the first footpoint emergence, i.e.,  698~s after the start of the series. The corresponding sequence from where the left panel is extracted can be found in a movie ($\it{v10}$) in the online edition.}
\label{bipole}
\end{figure}

Emergence event $\ell \it{10}$ starts with a host granule of 3\farcs3 diameter, as shown in the first panel of  the time sequence in Fig.~\ref{contourl122}. The thresholded patch of 15~G (in yellow) covers the central part of the granule in the first frames. This threshold is chosen to highlight the  unipolar character -- positive in both $\ell \it{10}$ and $\it{v10}$ -- of the emerging structures. As the granule grows and starts showing dark patches and crinkles that will become intergranular lanes, the magnetic patches overlie these dark areas. At 319~s from the start of the series, the buoyancy braking darkening is clearly visible (signaled by an arrow). This darkening is related to low velocities toward the observer, ending with moderate downflows of 0.5~\kms. The unipolar magnetic concentrations with strengths larger than 45~G (red/blue for positive/negative polarities) which develop into the recently created intergranular lanes do not remain unipolar for more than 2 frames (1~min), appearing an opposite po\-la\-rity nearby (see third and fourth panels in Fig.~\ref{contourl122}). This central inter\-granular lane drifts as the  granule grows, and another dark spot appears at 725 s, when the granule finally splits into four, becoming a recurrent explo\-ding granule. The area covered by the 15~G contours ranges from 18\% to 32\% of the granular area. Unfortunately, as the $\ell \it{10}$ emergence is taken without $Q$ and $U$ Stokes profiles, we cannot establish whether or not these positive and negative patches are actually linked as loop footpoints. 
In emergence event $\it{v10}$ we study a granule of initial dia\-meter around 2\arcsec~(sequence shown in the online material, movie $\it{v10}$). At  231~s from the start of the series, the center of the gra\-nule presents a sig\-ni\-fi\-cant area covered by longitudinal magnetic fields larger than 15~G. At 565~s, a slightly darker spot arises in the granule, together with a longitudinal magnetic patch of value larger than 45~G. The 15~G contours start forming a trefoil shape, covering from 15\% to a peak of 42\% of the granular area. The time span of this structure is 600 s approximately. It covers a granule partially, until the moment it approaches an intergranular lane or a breaking granule spot, where the longitudinal magnetic field migrates and intensifies (from 45~G to more than 80~G). LOS-velocities are slightly negative (upflows) while the magnetic field at the center of the gra\-nule is emerging. Maximum values of the longitudinal magnetic field peak around 100~G, while the transverse field is about 250~G. The maximum value of the transverse field is reached on the loop described below. Note that it seems a high value, but noise is quite large \citep[see][]{MartinezPillet2011}, then it is thresholded to 180~G. A positive polarity emerges at the edge of a previously fragmented portion of the granule followed by a transverse field patch at 764 s, indicating a magnetic loop top, lasting for around 90~s before the se\-cond footpoint becomes visible in our data. This dipole presents the semi-major axis pa\-rallel to the intergranular lane. The total duration over which the whole loop, including its apex is visible, is 396~s, also shown in Fig.~\ref{bipole} $(right)$, where a time series of the images in the region outlined by the white rectangle is presented. Green contours outline the transverse field while the background image corresponds to the longitudinal magnetic field. Positive/negative polarities are represented in white/black. Footpoints reach a maximum separation of 1.5 Mm, and the separation velocity is 3~\kms, much larger than the gra\-nule expansion rate, bearing in mind that the velocity computed between any two points is usually larger than its average. LOS-velocities at the footpoints range from 0.5~\kms~at the emergence of the first footpoint to 1.5~\kms. $\it{v10}$ event estimated flux ranges between 0.4-2.0~$\times$ 10$^{18}$~Mx considering magnetic elements over 45~G.
\subsection{Flowmaps of magnetic and non-magnetic features}
\label{S:31}
\begin{figure*}
\centering
\includegraphics[angle=-90,width=0.65\linewidth]{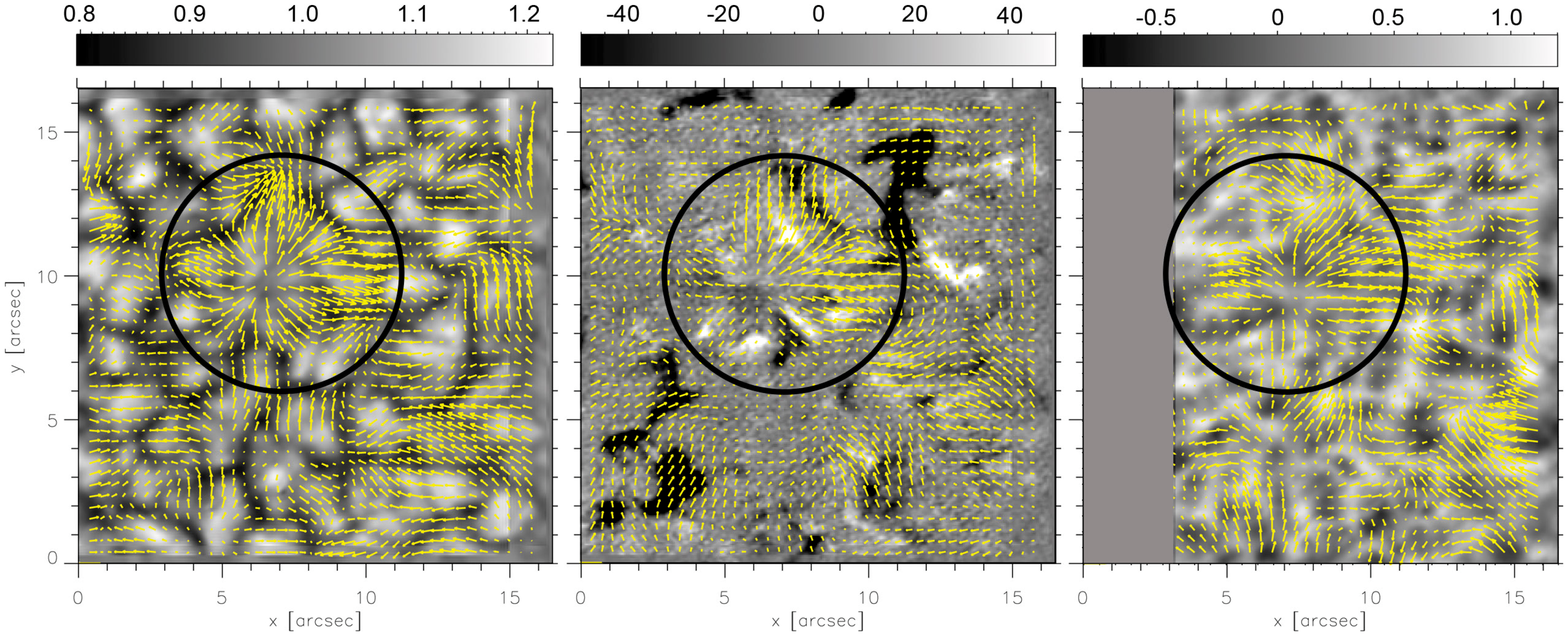}
\caption{{\it From left to right}: flow maps of the continuum, COG longitudinal field and LOS-velocity of the $\ell \it{10}$ emergence. Background images are average images of the following quantities: average normalized continuum image; longitudinal magnetic field, clipped to $\pm$ 50~G; and LOS-velocity (\kms~). The full FOV is 16\farcs5$\times$16\farcs5. Separation length between tickmarks is equivalent to a velocity magnitude of 1.5~\kms. A circle of diameter 8\farcs3  (same as in Fig.~\ref{contourl122}) is overplotted for comparison. See text for more details.}
\label{flowmapsl12}
\end{figure*}

\begin{figure*}
\centering
\includegraphics[angle=-90,width=0.65\linewidth]{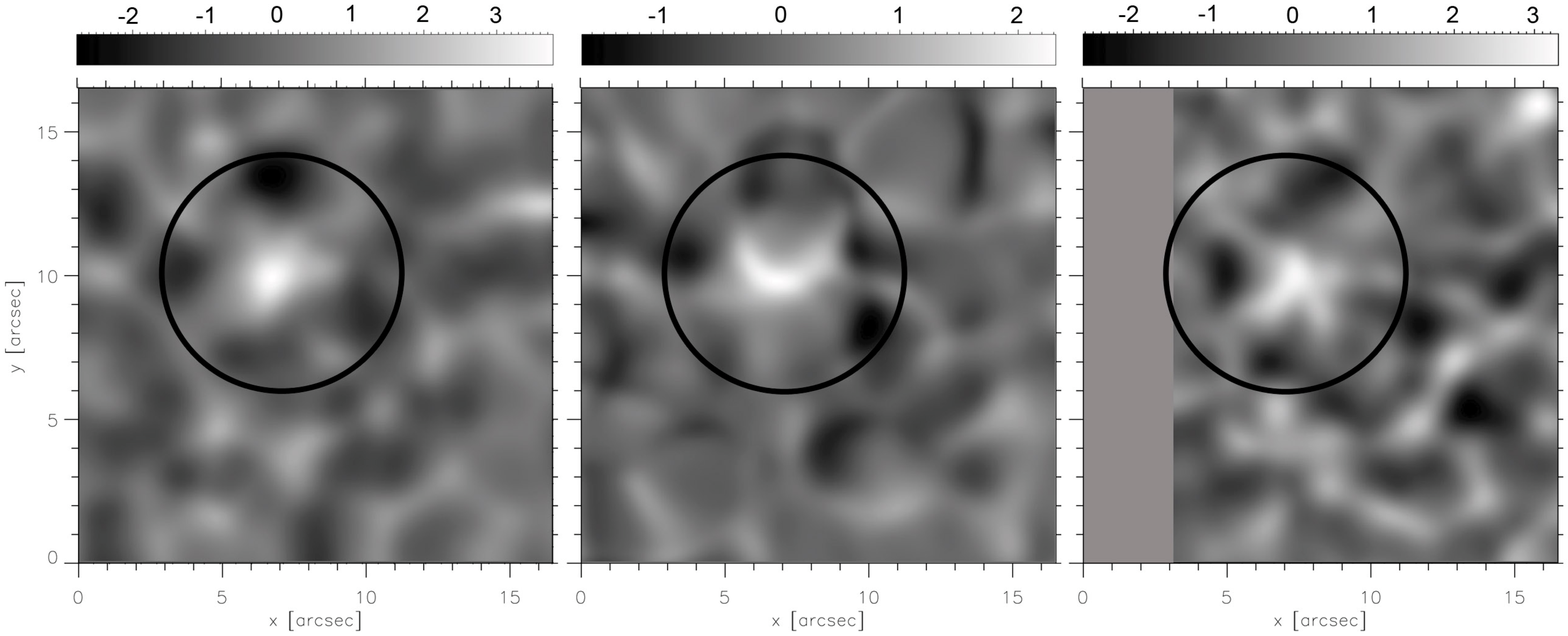}
\caption{{\it From left to right}: continuum, longitudinal field and LOS-velocity divergences (resulting from the corresponding flow maps in Fig.~\ref{flowmapsl12}) of $\ell \it{10}$ emergence. Bars indicate the divergence value in units of 10$^{-3}$ s$^{-1}$. The same circle of Fig.~\ref{contourl122} and \ref{flowmapsl12} is overplotted for comparison.}
\label{divl12}
\end{figure*}

In this study, the Local Correlation Tracking \citep[LCT,][]{November1988} method implemented by \citet{Molowny1994} is performed using a correlation window with {\it FWHM} of 1\farcs5 (28 IMaX pixels) averaged over the duration of the events (17 min for $\ell \it{10}$ and 21 min for $\it{v10}$). Flowmaps for continuum, the center-of-gravity longitudinal field and LOS-velocity are computed.  The time average LOS-velocity peaks between  ~~~-2.5 and 2~\kms.  Figure~\ref{flowmapsl12} displays the results for emergence event $\ell \it{10}$. Mean va\-lues of horizontal velocities in the continuum, longitudinal field and LOS-velocity maps are 0.63, 0.49 and 0.54~\kms, respectively. Velocities obtained with LCT are lower than expansion velocities because of the long duration of the series and the size of the {\it FWHM}. In addition, the flow pattern is remarkably similar in all the three panels of Fig.~\ref{flowmapsl12}. COG magnetic field flow maps show exactly the same pattern as the inversion magnetic field flow maps (not shown). The gray box in the last panel corresponds to a more stringent clipping of the inverted data (with respect to the non-inverted) to avoid influence from the apodised FOV, remnant of the reduction process. A circle of dia\-me\-ter 8\farcs3 is overlaid as visual indicator enclosing the exploding granule and its fragments. The divergence maps resulting from the LCT computation are shown in Fig.~\ref{divl12}. Maxima are found for each panel and averaging the position, we find an accurate center of the mesogranular flow. The coordinates averaged are x=6\farcs56$\pm $0\farcs30, y=9\farcs84$\pm$0\farcs11. Divergences peak around 3 $\times$ 10$^{-3}$ s$^{-1}$, which is a moderately high value for a 17-min average. Results for emergence event  $\it{v10}$ are comparable (average velocity va\-lues for images of continuum, longitudinal field, transverse field and Doppler velocity are 0.60, 0.47, 0.44 and 0.48~\kms) and hence not shown in this work. 
\subsection{Simulations}
\label{S:33}
\begin{figure}
\centering
\includegraphics[width=1.0\linewidth]{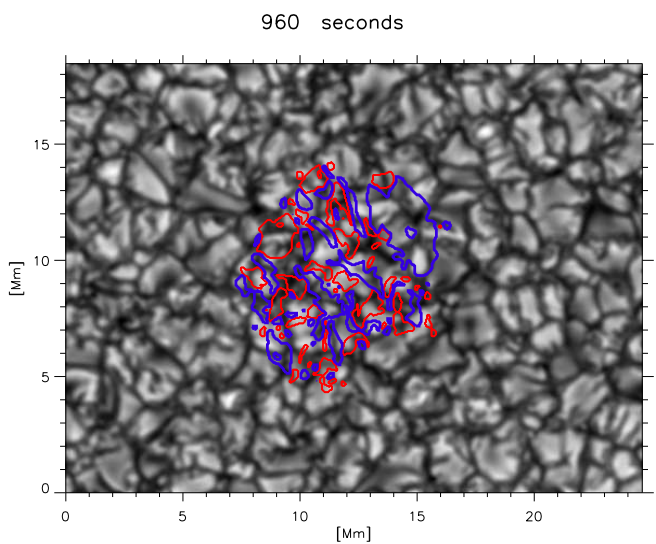}
\caption{Synthetic continuum image over-plotted with contours of the longitudinal component of the magnetic field (at $\tau_{500}$=0.1) coloured in red and blue, denoting 15 G and -15 G, respectively. The temporal evolution is displayed in a movie available in the online edition.}
\label{cheung_eps}
\end{figure}

\begin{figure*}
\centering
\includegraphics[angle=90,width=0.63\linewidth]{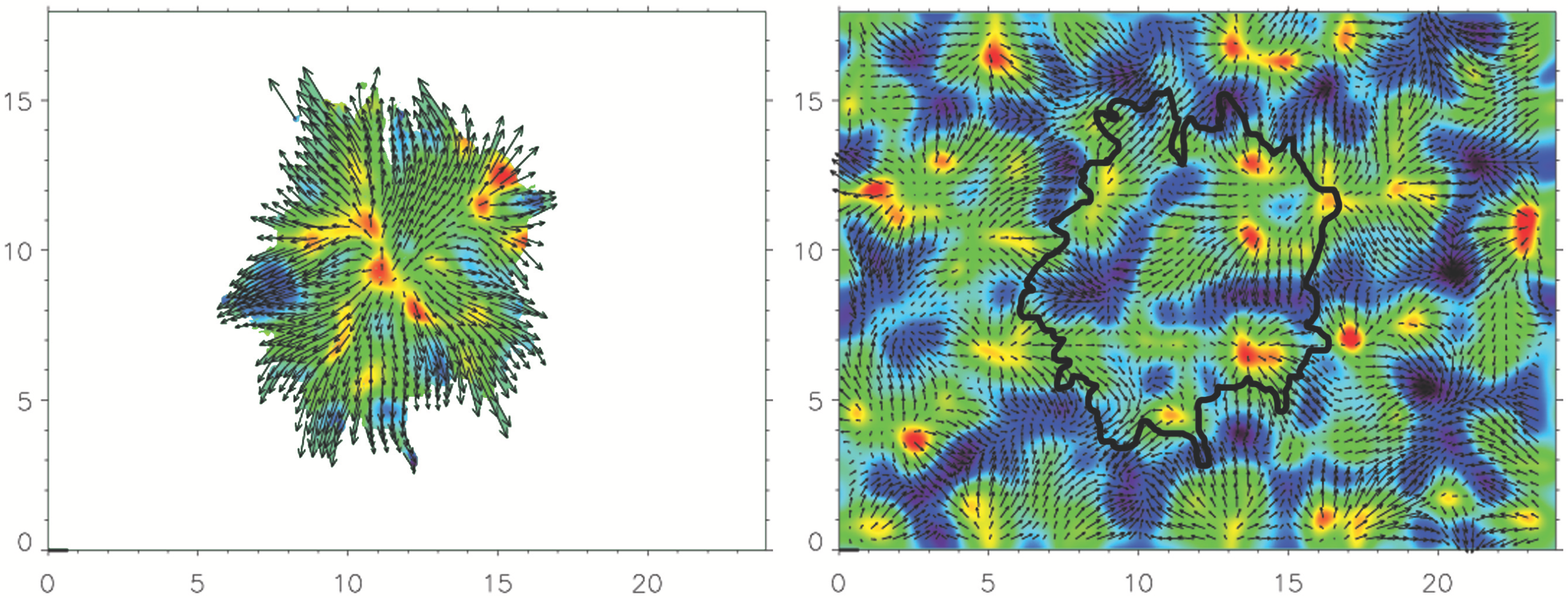}
\caption{LCT flow maps of the longitudinal magnetogram of the smoothed simulation (left) and its corresponding continuum image (right). Units in axes are Mm. Separation length between tickmarks is equivalent to a velocity magnitude of 1.1~\kms. The background image represents the normalized divergence field in every case, with positive values (in red), marking clearly the emergence areas, and negative divergences (in blue), featuring long-duration lanes and probably mesogranular lanes. Black contour delimits the magnetic field over the divergence in the right panel.}
\label{flowcheung}
\end{figure*}

A simulation performed by \citet{Cheung2008} is used for comparison. Carried out with the MURaM code \citep{Vogler2005}, which solves the radiative MHD equations on a Cartesian grid, this numerical experiment places a rising fluxtube with a twist parameter $\lambda$ =0.25 at  $z=-$3.9 Mm. The simulation domain has horizontal dimensions of 24 Mm $\times$ 18 Mm and height of 5.76 Mm, with a total magnetic flux of 10$^{20}$~Mx. A pixel of the simulation covers 25 km $\times$ 25 km or equivalently, 0\farcs033 px$^{-1}$, smaller than IMaX sampling, which is 0\farcs055 px$^{-1}$. The flux tube placed at the bottom of the simulation box expands and rises and becomes turbulent. The continuum images and the corresponding synthetic magnetograms extracted at $\tau_{500}$=0.1 are degraded by smoothing with a 0\farcs17 wide boxcar window, equivalent to 3 IMaX pixels.\\
 A movie of the simulation is available in the online material. This movie shows a continuum image with red and blue contours representing longitudinal fields larger than 15~G and lower than -15~G respectively. Maximum and minimum of the simulation are 220~G and~~-350~G. Similarly to the observed cases, some large unipolar blobs covering granules appear, and also smaller mixed polarities while the region evolves. Contrary to the observed cases, the longitudinal magnetic field seems to evolve more independently over the granules' top than in the analyzed events, where granules develop and magnetic flux is appa\-rently dragged by the plasma. Figure~\ref{cheung_eps} shows a  frame of the synthetic continuum overplotted with the aforementioned contours. The magnetic concentration centered in~~(14, 12) Mm is measured to estimate the expansion velocity in the simulation.\\
To analyze the behaviour of the simulation features with res\-pe\-ct to observation, we apply the same methods for expansion velocity calculation and LCT.  We computed the expansion velocity by means of effective radius measurement over time. The simulation holds the magnetic flux conservation and therefore, the magnetic patch area grows linearly with time. A negative patch covering some granules  actually expands at a rate of  1.2~\kms, considering the features with a pixel averaged field strength above the threshold of  -15~G. LCT ({\it FWHM} 1\arcsec~in this case) is applied also to the simulation. Fig.~\ref{flowcheung} displays the flow maps of the synthetic longitudinal magnetogram (left) and the continuum intensity (right). The ave\-ra\-ge longitudinal magnetogram is contoured over the continuum flow map (thick black line). While in the observed data a smaller window size does not provide more information, as tracked features are slower, in the simulation it highlights different features such as the divergence centers. Positive divergence centers are clearly spotted in red, while blue lanes indicate fragments of long-duration inter\-gra\-nu\-lar lanes, likely to be tracing mesogranular lanes (right panel). Most divergence features are well reproduced in both flow maps, e.g., strong negative divergence patch at coordinates (9, 9)~Mm, although more marked and the overall expanding behaviour more obvious in the left panel in Fig.~\ref{flowcheung}, i.e., when tracking magnetic features.
\section{Discussion and conclusions}
\label{S:4}
Thanks to the high resolution and cadence of $\textsc{\large Sunrise}$/IMaX, we obtained co-spatial and co-temporal magnetic field and continuum images, which allows us to study these events. For the first time we found magnetic loops embedded in the fragments of large exploding granules, which are covered by magnetic longitudinal field patches. The origin of these magnetic structures is still uncertain, although the similarity of the observed features with the simulations of \citet{Cheung2008} suggests that they originate by a turbulent rising flux tube. The emerging magnetic flux features are advected by the exploding granule. In this scenario the magnetic flux concentrations act as tracers. These two cases are the results of IMaX for two data sets of June 10, 2011, and more of them can be identified in the future, allowing us to study statistically their spatial and temporal occurrence.\\
In both emergence events, underlying granules expand at a rate of 0.95~\kms~and the magnetic patches at 0.65~\kms. Average horizontal velocities are about 0.60~\kms~for continuum and 0.46~\kms~for magnetic fields, in agreement with the typical granular expansion velocity estimated by LCT of 0.5-1.0~\kms~\citep{November1988}. The velocities are lower when magnetic structures are involved. This lower mobility of the magnetic features seems to be real and not a consequence of tracking faint features. This phenomenon was observed by \citet{Manso2011} as well. A plausible explanation can be that the magnetic field is measured higher up in the atmosphere than the continuum, leading to slightly lower velocities.\\
The emergence event $\it{v10}$ seems to be a long-lived recurrent explo\-ding granule \citep[$\it{active}$ granule,][]{Oda1984}. \citet[][]{Roudier2004} observed trees of fragmenting granules tracked over hours but no magnetic field information was available. Both cases presented here show magnetic field over gra\-nules, in agreement with \citet{Pontieu2002, Orozco2008}. Emergence event $\it{v10}$ occurs close to a network element. The recurrent exploding granule and associated emergence event $\ell \it{10}$ also harbours a network element in its surroundings. These facts could signify that these mesogranular emergences are close to a supergranular boundary. \\
Simulations used for comparison show a resemblance to the emergence $\ell \it{10}$, as the positive and negative polarities become visible very close to each other, albeit the field emergence is by far faster in the supergranular simulation than in the observed data. Considering a synthetic magnetic field concentration, the expansion velocity is in agreement to \citet{Tortosa2009}, who report expansion values of 4~\kms~on their simulations. While in \citet{Cheung2008} supersonic downflows are revealed, pro\-ba\-bly related to convective collapses, we find no evidence of supersonic up- or downflows in Stokes V con\-tinuum using V5-6 data (event $\it{v10}$) similar to those found by \citet{Borrero2010}.\\
Taking advantage of the $\it{v10}$ emergence that presents transverse field, we found a magnetic loop emerging in the FOV, whose asso\-cia\-ted footpoints separate from each other with a velocity of  $\sim$3~\kms, in agreement with \citet{Martinez2009} and \citet{Vargas2011}. The footpoints present downflows, indicate mass draining along the legs of the loops, making their ascent to less dense layers possible. As only one footpoint and the loop apex appear first, this may indicate the emergence of an inclined loop, or significant differences between the field strength of the two footpoints. As a general fact, the transverse field is almost negligible in the exploding granule, while longitudinal field covers portions of the granule before drifting to the intergranular lane where it intensifies.\\
\begin{acknowledgements}
The authors would like to thank to Mark Cheung for his simulations. JP acknowledges funding from the Spanish FPI grant BES-2007-16584. The German contribution to $\textsc{\large Sunrise}$ is funded by the Bundesministerium f\"ur Wirtschaft und Technologie through Deutsches Zentrum f\"ur Luft- und Raumfahrt e.V. (DLR), Grant No. 50 OU 0401, and by the Innovationsfond of the President of the Max Planck Society (MPG). The Spanish contribution has been funded by the Spanish MICINN under projects ESP2006-13030-C06 and AYA2009-14105-C06 (including European FEDER funds). The HAO contribution was partly funded through NASA grant number NNX08AH38G. This work has been partly supported by the WCU grant (No R31-10016) funded by the Korean Ministry of Education, Science \& Technology.
\end{acknowledgements}

\bibliographystyle{aa}

\end{document}